\documentclass[aps,twocolumn,prl,amssymb,amsmath, bm, floatfix,superscriptaddress,10pt]{revtex4}

\newcommand \beq{\begin{eqnarray}}
\newcommand \eeq{\end{eqnarray}}

\usepackage{graphicx}
\usepackage{mathptmx}
\begin{document}
\title{Towards a metallurgy of neutron star crusts}
\author{D. Kobyakov}
\affiliation{Department of Physics, Ume{\aa} University, 901 87 Ume{\aa}, Sweden}
\affiliation{Radiophysics Department, Nizhny Novgorod State University, Gagarin Ave. 23, 603950 Nizhny Novgorod, Russia}
\author{C. J. Pethick}
\affiliation{The Niels Bohr International Academy, The Niels Bohr Institute, University of Copenhagen, Blegdamsvej 17, DK-2100 Copenhagen \O, Denmark}
\affiliation{NORDITA, KTH Royal Institute of Technology and Stockholm University, Roslagstullsbacken 23, SE-10691 Stockholm, Sweden}
\begin{abstract}
In the standard picture of the crust of a neutron star, matter there is simple: a body-centered-cubic (bcc) lattice of nuclei immersed in an essentially uniform electron gas.  We show that at densities above that for neutron drip ($\sim 4 \times 10^{11}$) g cm$^{-3}$ or roughly one thousandth of nuclear matter density, the interstitial neutrons give rise to an attractive interaction between nuclei that renders the lattice unstable.  We argue that the likely equilibrium structure is similar to that in displacive ferroelectric materials such as BaTiO$_3$. As a consequence, properties of matter in the inner crust are expected to be much richer than previously appreciated and we mention consequences for observable neutron star properties.

\end{abstract}

\maketitle

Many technologically important properties of terrestrial metals are governed by the fact that these materials exhibit a variety of crystal structures. Pure metals have many different phases \cite{grimvall}. For alloys, even more possibilities exist, and these have far-reaching implications: e.g.,  the strength of steels is determined to a high degree by the existence of different crystal structures.   Here we consider matter in the outer parts of a neutron star (its crust), which is important for interpreting observations of neutron stars even though it comprises only a small fraction of the total mass of the star.  In the traditional view,  this matter is simple, because correlations between electrons, which are crucial for terrestrial matter, play little role. However, at densities above one thousandth of nuclear density, matter consists of a crystal lattice of atomic nuclei permeated by neutrons \cite{haensel}.  The neutrons behave like a second component in a binary alloy, and we argue that, as  a consequence, the properties of matter are more similar to those of terrestrial solids than has been previously appreciated.    Specifically, the neutrons give rise to an attractive interaction between nuclei which makes the lattice unstable to clumping of nuclei in a manner similar to the formation of inhomogeneous regions in metallic alloys (spinodal decomposition) \cite{sedrakian}.   While the attraction is insufficient to make matter unstable to long-wavelength distortions, it can destabilize matter at finite wavelengths where the effective interaction between nuclei due to their electrical charges is reduced.   We describe a number of possible consequences for observable properties of neutron stars.

To set the scene we consider the condition for thermodynamic stability of the system of nuclei immersed in a sea of neutrons, together with a background of electrons whose average density is the same as that of the protons to ensure electrical neutrality.  The system may thus be regarded as having two components,  the neutrons (both those in nuclei and the interstitial ones), and the charged particles.   For most of the life of a neutron star, the temperature is so low that thermal effects may be neglected.  In that case, the condition for stability is that the second order change in the energy density be positive for neutron and proton densities slightly different from the initial ones  (which are determined by the condition that matter is in equilibrium to weak interaction processes), i.e.
\beq
\delta^2 E=\frac12 \sum_{i,j}\delta n_i \delta n_jE_{ij}  >0,
\eeq
where $E_{ij}  ={\partial^2E}/{\partial n_i \partial n_j}$ \cite{landau}.  The species labels $i$ and $j$ here refer to neutrons ($n$) and protons ($p$) and the $n_i$ are particle number densities.  For electrically neutral matter the electron density $n_e$ is equal to the proton density and is therefore not an independent variable.  The quantity $E(n_n,n_p)$ is the energy of the system, including electrons, per unit volume.  Sufficient conditions for stability are that one of the diagonal elements of $E_{ij}$ be positive and that the determinant of the matrix, $E_{pp}E_{nn}-E_{np}^2$, be positive. Since $E_{nn}>0$, the latter condition may be rewritten as
\beq
E_{pp}-\frac{E_{np}^2}{E_{nn}}=\left.\frac{\partial \mu_p}{\partial n_p}\right|_{\mu_n}>0,
\label{stab}
\eeq
where the $\mu_i$ are chemical potentials.
Equation (\ref{stab}) has the physical interpretation that the effective proton--proton interaction be positive: the first term is usually referred to as the direct interaction, because it represents the interaction when the neutron density is held fixed, while the second term, the so-called induced interaction, is the contribution to the interaction due to the change in the neutron density.   The second term, which is attractive, is similar to the phonon-induced interaction between electrons that is responsible for conventional superconductivity in terrestrial metals, except that in a neutron star the exchanged phonon is a neutron density fluctuation.

\begin{figure}
\includegraphics[width=3.3in]{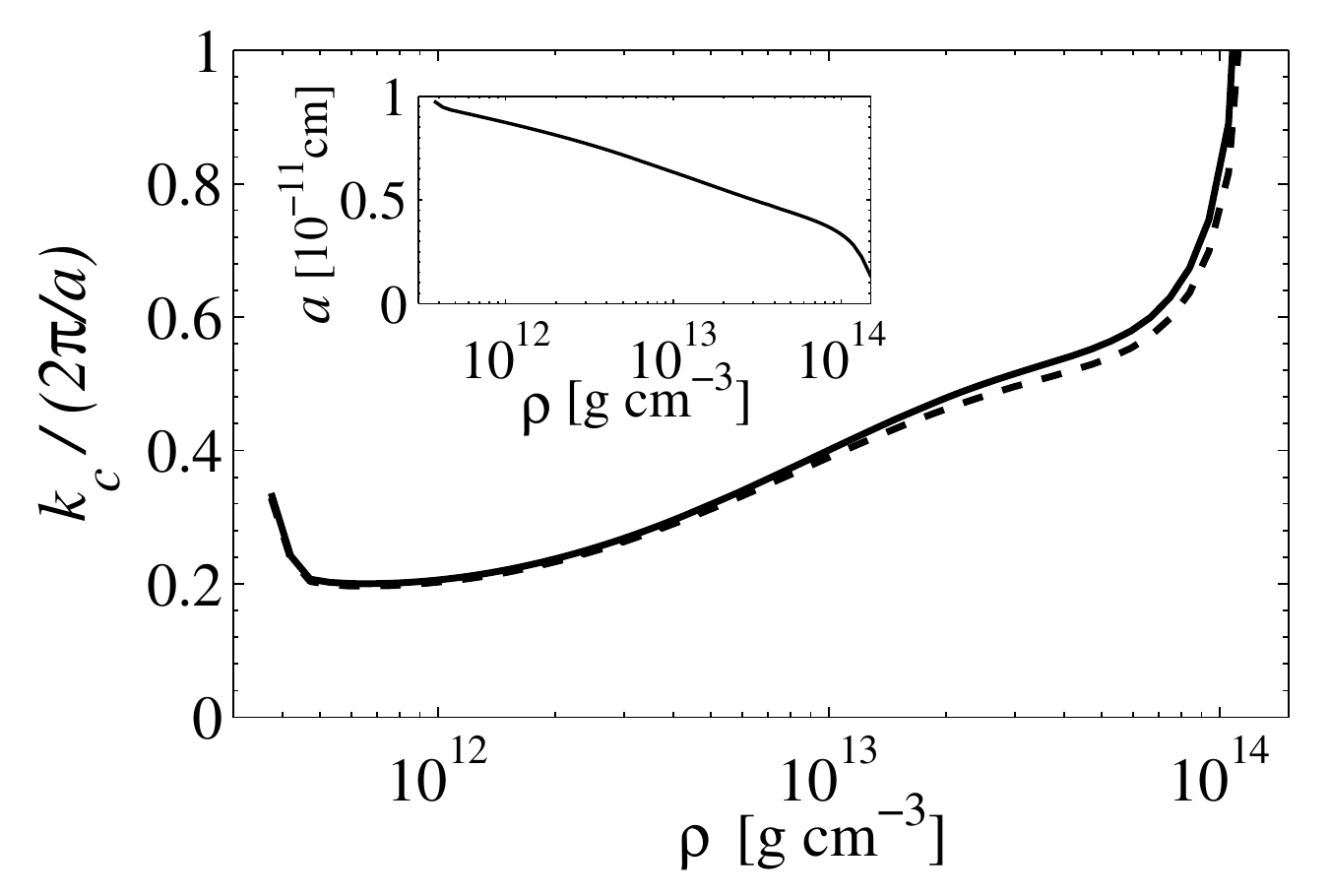}
\caption{Wave number for the onset of instability compared with the wavenumber $2\pi/a$, which gives the extent of the first Brillouin zone along a cubic axis, as a function of density. The solid line shows results when the shear elastic constants are included (Eq. (7)) while the dashed line shows the results when they are neglected (Eq. (6)). The inset shows the lattice spacing as a function of density.}
\end{figure}
\begin{figure}
\includegraphics[width=3.5in]{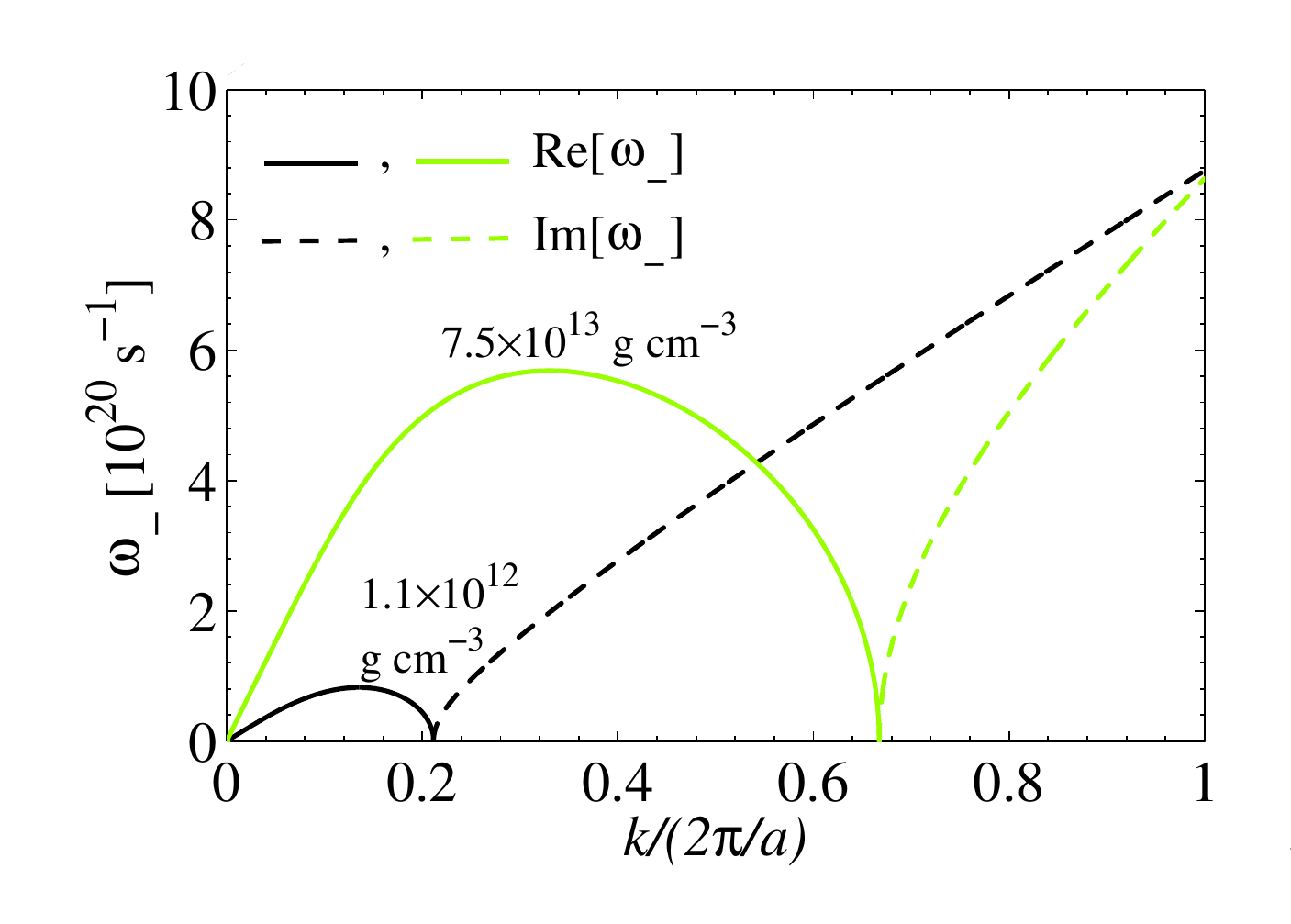}
\vspace{-0em}
\caption{(Color online) Frequency  $\omega_-=kv_-$ for the mode in which neutrons and protons move in phase with each other.  When $\omega_-$ is real, its value gives the oscillation frequency (solid lines) while when it is imaginary, the mode is unstable with a growth rate  ${\rm Im}(\omega_-)$ (dashed lines).   When the periodicity of the system is properly taken into account, the dispersion relation will have zero slope at the zone boundary ($k=2\pi/a$) for $k$ along one of the cubic axes. }
\end{figure}
\begin{figure}
\includegraphics[width=3.5in]{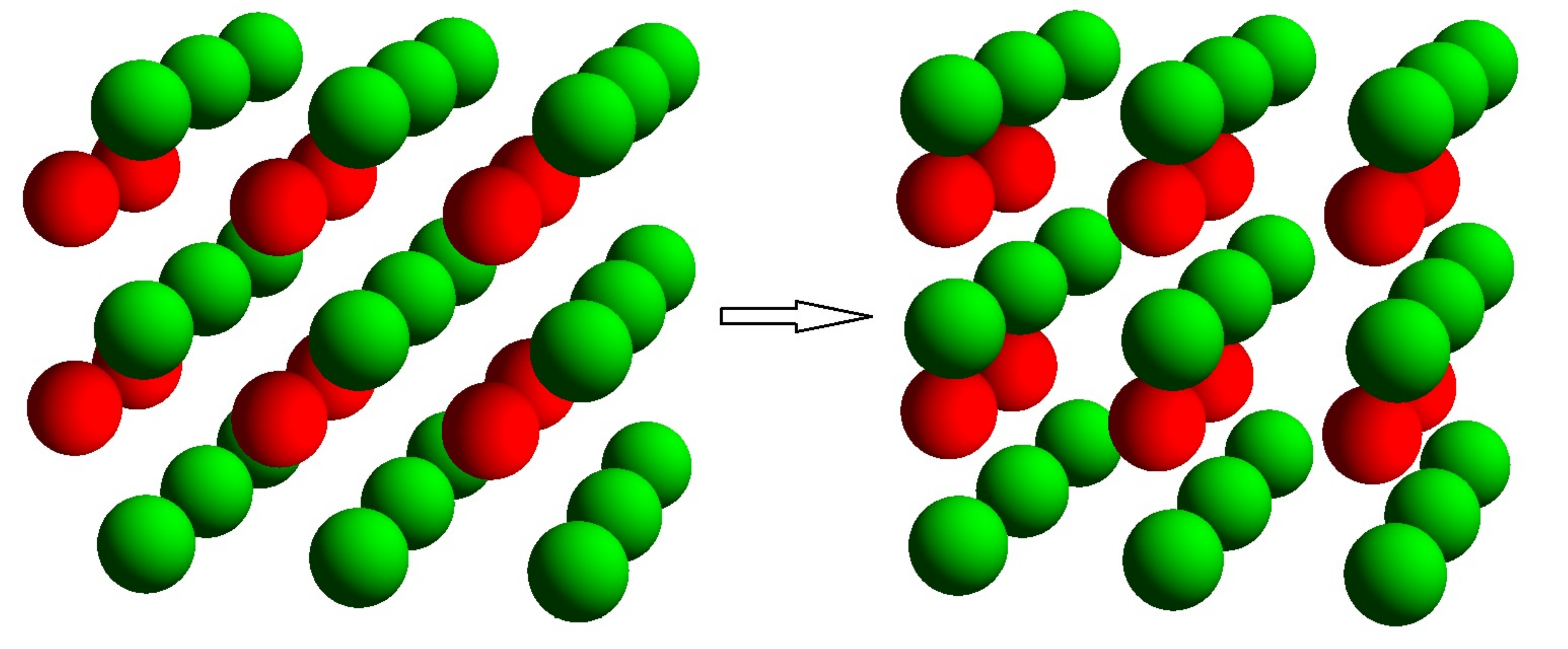}
\vspace{-0em}
\caption{(Color online) Schematic picture of the undistorted bcc lattice (left) and the structure in which the central atoms in the cell are displaced to the right along one of the cubic axes (right), thereby creating a superlattice structure with two atoms per cell. While the nuclei at the centers of cubic cells are identical with those at the corners, for clarity we show nuclei at the centers of unit cells of the undistorted structure as red (dark grey) and those at the corners as green (light grey).}
\end{figure}

Calculation of the derivatives $E_{ij}$ is a challenge because the system has two phases, the nuclei and the neutrons between nuclei.  Consequently the derivatives are not simply related to those of bulk nuclear matter.  In order to obtain thermodynamically consistent results,  we have recently evaluated them from Lattimer and Swesty's microscopic calculations of the properties of dense matter\cite{lattimer, kobyakov}, and thereby take into account the fact that when neutrons are added to the system, they are distributed between the two phases. Our results show that matter is stable for bulk disturbances since the magnitude of the induced interaction is typically no more than 0.2-0.3 times the direct interaction \cite{earlierwork}.  The stability  is due to the
fact that when the proton density is increased, the electron density must also increase and the effective proton--proton interaction is dominated by the energy required to compress the electrons.

The situation is different at finite wavelengths, or nonzero wave number $k$.  The most important contribution to the proton--proton \cite{ppinteraction} interaction is due to the Coulomb interaction.  This is screened by the electrons and is given by \cite{pines}
\beq
V(k)=\frac{V_0(k)}{1+V_0(k) \chi_e(k)},
\eeq
where $V_0(k)=4\pi e^2/k^2$ is the bare Coulomb interaction and $\chi_e(k)$ is the electron density-density response function.  For small $k$, $\chi_e(k)$ tends to $\partial n_e/\partial \mu_e \equiv 1/E_{ee}$, and therefore $V(k)$ tends to $\partial \mu_e/\partial n_e$.  Inserting standard results for a relativistic gas, one finds that for $k$ much less than the Fermi wave number, the effective proton--proton interaction is given by \cite[Ch. 2]{haensel}
\beq
V(k)=\frac{V(0)}{1+k^2/k_{\rm TF}^2},
\label{screenedcoulomb}
\eeq
where the Thomas--Fermi screening wave number is given by $k_{\rm TF}^2=(4\alpha/\pi)k_e^2$, $\alpha=e^2/\hbar c$ being the fine structure constant and $k_e$ the electron Fermi wave number.  To approximate the effective proton--proton interaction for nonzero $k$ we take the screened Coulomb interaction (\ref{screenedcoulomb}) plus a constant term whose magnitude is such as to ensure the correct $k=0$ limit:
\beq
E_{pp}(k)=E_{pp}-\frac{\partial \mu_e}{\partial n_e}\frac{k^2}{k_{\rm FT}^2+k^2}.
\label{Epp(k)}
\eeq
This equation expresses the fact that it is less costly energetically to create proton density modulations because electrons do not follow the protons if the wave number is comparable to or greater than $k_{\rm FT}$.

To investigate stability at nonzero $k$ we first assume that the instability condition is a direct generalization of Eq.\ (\ref{stab}) to this case.  We expect the $k$-dependence of $E_{nn}$ and $E_{np}$ to be on wave number scales of order $2\pi/a$, the extent of the first Brillouin zone,  and therefore we replace these parameters by their $k=0$ values.
Here $a$ is the distance between nuclei at the corners of a cubic cell.
The smallest unstable wave number $k_c$ is thus given by
\beq
\frac{k_c^2}{k_{\rm FT}^2}=\left[\left(  \frac{E_{np}^2}{E_{nn}}-E_{pp}\right) \frac1{E_{ee}}+1 \right]^{-1}-1,
\eeq
which reduces to $E_{nn}E_{pp}/E_{np}^2-1$ for $E_{ee}=E_{pp}$.
In Fig.\ 1 we show the calculated values of $k_c/(2\pi/a)$ as a function of density.  The predicted values of $k_c$ are well below $2\pi/a$, so our neglect of the $k$-dependence of the parameters is  a good first approximation.

Using the formalism of Ref.\ \cite{kobyakov}, we have calculated the angular frequency and the growth rates of the softest mode.  Since the effective neutron--proton interaction $E_{np}$ is negative,  the proton and neutron density variations are in phase for this mode, which corresponds to the one with frequency $kv_-$ in the notation of that paper.  Results are shown in Fig. 2 for two densities.  While the value of the critical wavenumber does not depend on the value of the neutron superfluid density, the frequencies and growth rates do, and for our calculations we have assumed that the number density of superfluid neutrons is equal to the density of neutrons outside nuclei.   From these results, one sees that the growth times of unstable modes are extremely short on the scale of the lifetime of a neutron star, and this conclusion is unaffected by uncertainties in the neutron superfluid density, which calculations of Chamel  indicate could be an order of magnitude smaller than the value we have taken \cite{chamel2012}.

So far we have assumed that the energy per unit volume depends only on the densities of neutrons, protons, and electrons; thus effects due to distortion of the crystal from its original cubic form have not been taken into account.  In the absence of interstitial neutrons, for a crystal with cubic symmetry there are three independent elastic constants, which are conveniently taken to be the bulk modulus $B=(c_{11}+2c_{12})/3$, the modulus $c_{44}$ that describes a shearing perpendicular to one of the cubic axis, and the quantity $c_{11}-c_{12}$, which describes response to a volume-conserving distortion with extension along one cubic axis and compression along one of the other axes.  Here we use the standard notation for elastic constants \cite{elasticconstants}.
For bulk stability, deformation of the crystal, $B$, $c_{11}-c_{12}$, and $c_{44}$ must all be greater than zero.  For a bcc Coulomb lattice, the shear elastic constants are dominated by the static lattice contribution and $c_{11}-c_{12}  \approx 0.10 n_i Z^2e^2/a$ and  $c_{44} \approx 0.37 n_iZ^2e^2/a$ \cite{fuchs}, where $n_i=n_e/Z=2/a^3$ is the density of nuclei.  For an isotropic solid the ratio $(c_{11}-c_{12})/2c_{44}$ is unity, while for a Coulomb crystal it is $\sim 0.13$. Crustal material is therefore unusually anisotropic, and similar to lithium and plutonium by this measure.  For this case, the wave vectors of the most unstable modes associated with a density fluctuation lie along the cubic axes, as Cahn showed in his classic analysis of spinodal instability, in which a binary alloy forms regions with different composition \cite{cahn}. The phase transition in the neutron star case differs from the usual spinodal instability in that it occurs at a finite wavelength.  The instability occurs first when $c_{11}=B +2(c_{11}-c_{12})/3=0$, which is more restrictive than the condition $B=0$ for a crystal with no rigidity to shear.  When interstitial neutrons are included, the instability condition is essentially the same, except that the bulk modulus is replaced by the effective proton--proton interaction $E_{pp}(k)-E_{np}^2/E_{nn}$ times $n_p^2$.  The critical wave number is thus given by
\beq
E_{pp}(k_c)-\frac{E_{np}^2}{E_{nn}} +\frac23\frac{(c_{11}-c_{12})}{n_p^2}=0.
\eeq
The solid line in Fig.\ 1 shows that the results of this calculation, and one sees that the inclusion of the shear elastic constants has little effect on the critical wave number. This is because contributions to the shear elastic constants are typically of order $Z^{2/3}\alpha \sim 0.1$ times the other terms in the instability condition.

On the basis of the instability analysis alone, it is impossible to predict the equilibrium structure, which depends on higher order terms in the expansion of the energy in powers of density deviations.  However, one would expect the actual structure of the metal to be modulated at wave vectors corresponding to the most unstable modes.  In our model, growth rates of modes increase with $k$ for $k>k_c$, Fig. 2, and the maximum growth rate would occur at the zone boundary, which is at a wave vector (1,0,0)\,$2\pi/a$ or some equivalent wave vector, where $a$ is the length of the edge of a cubic unit cell.  This corresponds to a doubling of the volume of the unit cell, which becomes a cube with two atoms per unit cell, the one in the middle of the cube being displaced from the center of the cube, as shown in Fig. 3.  The transition is thus analogous to that to a ferroelectric in materials such as BaTiO$_3$ \cite{cowley}, but in the neutron-star case there is no ferroelectricity, because the displaced nuclei at the centre of the cube are identical with the nuclei at the corners.  In terrestrial materials under pressure many examples of incommensurate phases transitions have been discovered \cite{mcmahon}, and further studies are necessary to determine if such a phase could occur in neutron star crusts.

In this article we have demonstrated that, because interstitial neutrons in the inner crust behave in a way similar to a second species in a metallic alloy, the physics of the crust is much richer than has been appreciated previously.
The phase transition we predict would have far-reaching consequences for a number of important properties of neutron star crusts,
and for interpreting observations from space-based instruments.
Thermodynamic and transport properties  of the crust are important in understanding cooling of neutron stars \cite{yakovlev}, and  glitch phenomena  \cite{Andersson2012, Chamel2013}.  The breaking strain of the crust, which is sensitive to phase transitions of the lattice, is important in a number of contexts.
For example, precursor flares prior to the short gamma-ray bursts \cite{Troja}
may be understood as being due to breaking of the crust
by tidal forces, or as the outcome of a resonance
of elastic modes
\cite{Tsang}. Future
gravitational wave measurements
are expected to detect mountains on accreting neutron
stars \cite{HorowitzKadau}, and the maximum height of the mountains depends on the breaking strain. From experience with terrestrial solids, one knows that the breaking strain of real materials is very different from, and often larger than, that for a perfect lattice, e.g., in martensitic steels.

With the current rapid advances in computational methods in nuclear
physics, as well as the increasing amount of experimental data
on neutron-rich nuclei that is becoming available, the way is
now open to take a fresh look at the properties of
matter in the inner crust.  One aspect of the problem that we have not addressed in this article is the role of impurity nuclei.  In terrestrial materials, these have an enormous influence on physical properties and their role in neutron stars needs to be reinvestigated.

We are very grateful to G\"{o}ran Grimvall for giving us an authoritative introduction to lattice instabilities in metals and to Pascale Deen, Maxim Mostovoy, David Nelson and Bertil Sundquist for helpful discussions.  CJP thanks Armen Sedrakian for drawing his attention to the possible role of interstitial neutrons in destabilizing matter in the crust of a neutron star. The work of DK was supported in part by the J. C. Kempe Memorial Fund.


\begin{thebibliography}
\bibitem{ppinteraction} Since all protons are in nuclei, the nucleus--nucleus interaction is simply $Z^2$ times the proton--proton one.
\end{thebibliography}
\end{document}